\documentclass[aps,prd, amsmath,amssymb]{revtex4-2}

\usepackage{graphicx}
\usepackage{bm}

\begin{document}

\title{Influence of wind on a viscous liquid film flowing down a thread}
\author{Annette Cazaubiel}
\email{annetca@math.uio.no}

\affiliation{Department of Mathematics, Mechanics Division, University of Oslo, N-0851 Oslo, Norway}
 \author{Andreas Carlson}
 \email{acarlson@math.uio.no}

\affiliation{Department of Mathematics, Mechanics Division, University of Oslo, N-0851 Oslo, Norway}
\date{\today}
\begin{abstract}
We investigate experimentally the effect of wind on the dynamics of a viscous liquid film flowing down a thread. The liquid film is well-known to destabilize into an axisymmetric bead-like pattern in stagnant air. When the flowing film is exposed to side-wind, the beads are pushed downstream breaking the symmetry leading to a modification of the film flow. The viscous dissipation in the liquid film decreases with increasing side-wind, leading to a faster slide speed of the liquid beads. The inertia in the film is then increased, which can induce a regime shift, triggering an instability. When the film becomes unstable, it switches from a periodic pattern of equally sized liquid beads to an irregular film pattern where large beads collide with smaller ones.
\end{abstract}
\maketitle

\section{Introduction}
\label{intro}

A thin liquid film flowing down a thread will destabilize and form a wide range of interfacial patterns that can be grouped into different flow regimes \cite{Kliakhandler_2001}. Such dynamics are important to a wide range of applications, such as heat exchanger \cite{Zeng_2017}, gas absorption \cite{Chinju_2000} or desalination processes \cite{Sadeghpour_2019}, which can also include an air stream. However the effect of a side-wind on a film flow, which is reported in this article,  has to the best of our knowledge not yet been characterized. 
The film flow of a liquid around a cylinder without external air flow can display a rich dynamical behavior, liquid beads (or thought of as drops) slide at constant speed, solitary waves are formed on the film and the film flow becomes irregular \citep{Kliakhandler_2001,Duprat_2009,Ruyer-Quil_2008}. Quéré \cite{Quere_1990} made an experimental investigation of film flow along threads by drawing wires from liquid baths. He observed the formation of beads when the film thickness was larger than $r_t^3l_c^{-2}$, with $r_t$ the radius being of the thread and $l_c$ the capillary length. These observations and this criterion were confirmed theoretically \cite{Frenkel_1992}, assuming the thickness of the film is much smaller than the radius $r_t$ of the thread and neglecting the effects of inertia in the flow. Kalliadasis and Chang \cite{Kalliadasis_1994} developed then a quantitative theory, including an evolution equation by also using the lubrication approximation. By continuously wetting the thread from the top, Klikhandler \textit{et al.} \cite{Kliakhandler_2001} reached flow rates for which lubrication theory does no longer hold. They revealed that by changing the flow rate the liquid film could go from a periodic pattern of droplike beads to a regime where they are large and collide irregularly. Those regimes correspond to two different instability mechanisms competing in the system. The first one is the Rayleigh-Plateau instability driven by surface tension revealed experimentally by Plateau \cite{Plateau} and explained theoretically by Rayleigh \cite{Rayleigh}. The second one, the Kapitza instability, described and analyzed by Kapitza \cite{Kapitza}, occurs when a liquid film flows down an inclined wall, then inertia induces the formation of capillary wave trains. Duprat \textit{et al.} \cite{Duprat_2009} investigated the interplay between these two instabilities and completed the regime map of Kliakhandler \textit{et al.} \cite{Kliakhandler_2001}. These experimental results were in close agreement with the theoretical model developed by Ruyer-Quil \textit{et al.} \cite{Ruyer-Quil_2008} including inertia and streamwise viscous diffusion, obtained by combining a long-wave approximation with a weighted residual technique. More recently, the effect of the nozzle geometry on the flow has been investigated. Sadeghpour \textit{et al.} \cite{Sadeghpour_2017} showed experimentally that the flow characteristics could vary significantly only by changing the nozzle diameter and Ji \textit{et al.} \cite{Ji_2019} used a complete lubrication model to explain these experimental observations.

To better apply this fundamental phenomenon to industrial applications and to gain control over the flow, different variations over the problem of a liquid film flowing down a thread have been investigated. 
 Ding \textit{et al.} \cite{Ding_2014} showed that a radial electric field on a film of a perfectly conducting liquid could either enhance or impede the instability depending on the magnitude of different parameters. Due to their relevance in many industrial processes and in particular coating flows, thermocapillary effects on liquid film flows have been extensively studied, see for example \cite{Kabova_2012}, \cite{Liu_2018} and \cite{Ding_2018}. The stability of a liquid film around a rotating thread was analyzed by Liu \textit{et al.} \cite{Liu_2020}. Coating flows of non Newtonian fluids have also been examined, for instance it was shown that micelle solutions stabilize the film \cite{Boulogne_SM_2013} and it was established that polymer solutions decrease the growth rate of the instability \cite{Boulogne_2012}.  
Recently, Gabbard \textit{et al.} \cite{Gabbard_2021} observed that the film could lose its axisymmetry for certain values of the thread radius $r_t$ and the surface tension $\sigma$. They reported that the dynamics was more regular when the axisymmetry was broken. The first effect of a side-wind on a film flowing down a thread will be breaking  the symmetry due to the drag force.  Gilet \textit{et al.} \cite{Gilet_2010} have shown that when a drop sliding down a thread, going from a non-axisymmetrical geometry to an axisymmetrical geometry, its sliding velocity decreases significantly. It can therefore be expected that the wind will have a significant impact on the velocity of the sliding liquid beads in the film flow.
Furthermore, the effect of side-wind has been shown to be important and non-trivial in the dynamics of drop on threads. It was shown that  the drag of the air flow can induce oscillations and rotations of clamshell as well as barrel shaped drops on a thread \citep{Mullins_2005,Mullins_2006}. Bintein \textit{et al.} \cite{Bintein_2019} revealed that drops could be put into motion by their asymmetric wakes in presence of a strong enough side-wind. Lastly, if the drag force is strong enough, drops are detached from the thread or fiber \cite{Sahu_2013}. 
 
 In this paper we investigate experimentally how side-wind influences the beading patterns of a film flow around a thread when the Rayleigh-Plateau instability dominates. We analyze how the drag force pushes the liquid film downstream, breaking the symmetry of the system and  how it changes the flow.  Side-wind is shown to make the liquid beads slide up to twice faster, which can lead to a transition from a periodic to an irregular film flow regime.

\section{Experiments}
\subsection{Experimental setup}
\begin{figure}
\includegraphics[width=\textwidth]{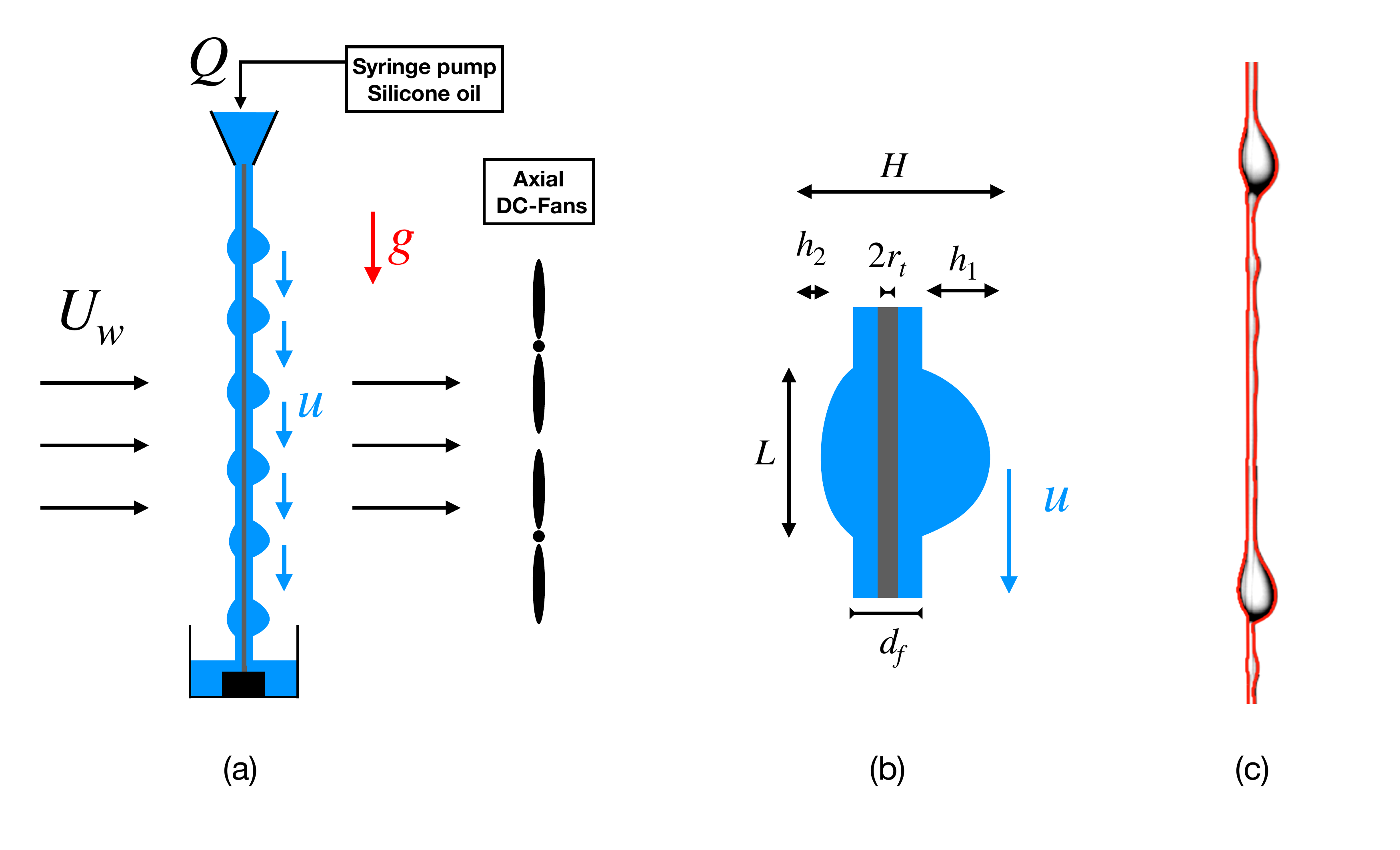}
   \caption{(a) Schematic description of the experimental setup. (b) Sketch of the fluid film and an asymmetrical bead described by its geometrical properties: $r_t$, the radius of the thread, $d_f$, the diameter of the film, $H$ and $L$, the width and the length of the bead, $U$ its sliding speed in the direction of gravity, $h_1$ and $h_2$ the downstream and upstream heights of the bead. (c) An experimental snapshot in time of the film, where the side-wind, coming from the left, is pushing the beads downstream, the geometrical properties of the film are deduced from the extracted profile represented in red on the picture.}
\label{setup}
\end{figure}

We study the flow of a film of silicon oil v10, v50 and v100 (with kinematic viscosity $\nu=$ [1$, 5$, 10$]\times 10^{-5}$ $\text{m}^2$/s, of density $\rho=$ [0.93$, 0.96$, 0.97]$\times 10^3$ kg/$\text{m}^3$ and of surface tension coefficient $\sigma=$ 20.1, 20.8, 20.9 mN/m, respectively) on a vertical Nylon line of radius 0.125 mm $\leq r_\text{t}\leq$0.3 mm. These oils perfectly wet the nylon thread. The fluid is pushed through a needle around the Nylon thread with a syringe pump (model Aladdin 1000 of World Precision Instruments) at a constant flow rate $Q$ (0.5 mL/min$\leq Q\leq$5mL/min). Air is surrounding the thread at room temperature \mbox{$T\approx$ 22 \textdegree C}. The Nylon threads are strongly tightened to avoid any wind-induced vibrations.

\begin{figure}
\includegraphics[width=0.9\textwidth]{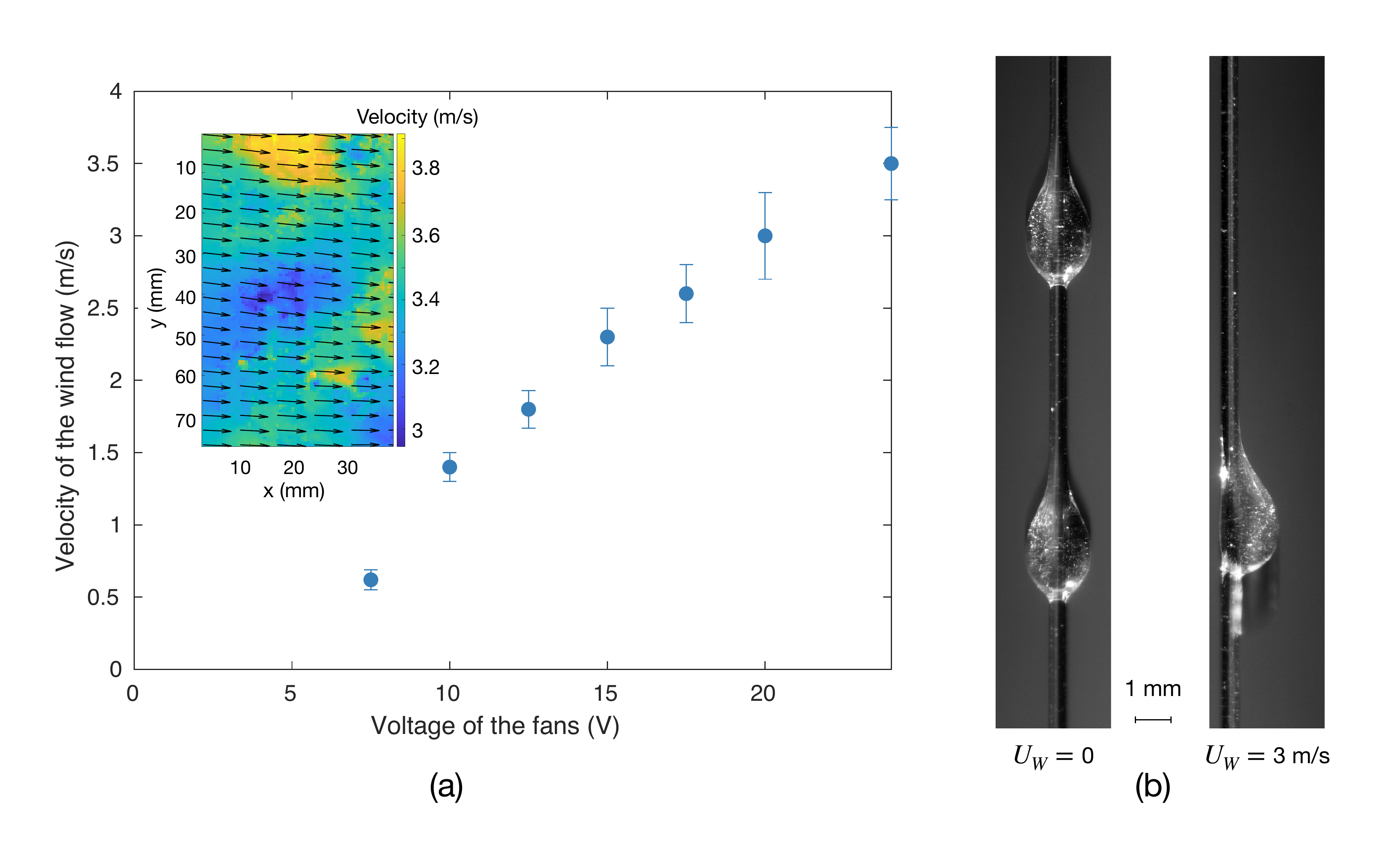}

   \caption{(a) Wind velocity created averaged over space and time as a function of the voltage imposed on the fans without contractions in the channel. Inset: Instantaneous wind field for the maximum fan speed. (b) Images of the liquid film seeded with tracer particles in the laser sheet for PIV-measurements of the flow, left: without wind, right: with a wind-speed $U_W=3$ m/s.}
\label{wind}
\end{figure}

The experimental setup is presented in Figure \ref{setup} (a), where the wind flow is generated in a channel of width 25 cm, height 60 cm and length 600 cm. 
The generated wind flow has been characterized through Particle Image Velocimetry (PIV) measurements using fog droplets of size 0.1 mm as tracers.  At one end of the channel, two axial fans of 225 mm diameter create a flow by sucking air.  We get an horizontal airflow between 0.5 and 3.5 m/s depending on the voltage imposed on the DC-fans as shown in Figure \ref{wind}  (a). An example of the instantaneous wind field for the maximum fan velocity is shown in Figure \ref{wind} (b). 
To reach higher velocities and observe the detachment of the liquid beads, a contracting section was placed in the channel to reduce its effective height to 15 cm allowing us to reach a wind speed up to 14 m/s. However, this leads to a reduced observation window of the thin film flow.

The dynamics of the liquid film was recorded by a high-speed camera (model FASTCAM Mini WX100 by Photron) with a macro-lens (Carl Zeiss Makro-Planar T* 2/100 by Nikon) at 250 frame per seconds. The images are then binarized and the profile of the liquid film is extracted as shown in Figure \ref{setup} (c). We call the largest structures formed `beads' (similar in form as drops). The geometrical properties of the sliding beads, defined in Figure \ref{setup} (b) are then extracted from the profile of the liquid film. 

In order to capture the dynamics of the flow within the film, we performed PIV-measurements in the following way. The silicone oil is seeded with silver coated hollow glass spheres (S-HGS by Dantec), of diameter 10 $\mu$m and of density 1.4$\times 10^{-3} kg/m^3$. A double pulse laser (ICE 450 by Quantel) creates a laser sheet in a plan parallel to the wind and containing the thread. The particles inside the liquid film are then filmed by our high-speed camera. Figure \ref{wind} (b) show typical images of the seeded film without wind (left panel) and with a wind speed $U_W=3$ m/s (right panel). The small white dots inside the film are the illuminated tracer particles. The vector field is then calculated by using the PIVlab algorithm of Matlab \cite{PIVlab}. The sliding velocity of the beads is then subtracted in order to see the vector field in the frame of reference of the liquid beads.

Figure \ref{picture} illustrates the liquid films shape for three different thread radii (a) with no wind, (b) with a wind speed  $U_w=1.4$ m/s and (c) with $U_w=3$ m/s. Without wind, we observe the formation of axisymmetric structures. The wind breaks the film symmetry, in particular the liquid beads are pushed on one side of the thread. 
\subsection{Parameters and non-dimensional numbers}

We use the following non-dimensional numbers to get insight on the physics of this flow phenomenon. To quantify the drag force and the effect of the side-wind on the system, we compute the Reynolds number of the liquid beads in the wind flow, $Re_W=HU_W/\nu_\text{air}$, where $H$ is the width of the bead (see Fig. \ref{setup}), $U_W$, the velocity of the air stream and $\nu_\text{air}$ the kinematic viscosity of air. For most experiments, $0<Re_W<600$. To study liquid bead detachment, the Reynolds number of the wind can reach up to $Re_W=2100$. The change of shape of the liquid bead is estimated by the asymmetry coefficient: $a=(h_1-h_2)/(h_1+h_2)$, with $h_1$ and $h_2$ the heights downstream and upstream of the bead with respect to the wind direction, respectively. To predict the onset of detachment, we introduce the Bond number, ($Bo$) to balance capillary adhesion to the thread with the drag force, $Bo=F_D/(4\pi \sigma r_t)$, with $F_D=1/2 C_D \pi H^2 \rho_\text{air} U_W^2$ the drag force, where $C_D$ is the drag coefficient. We assume that the drag coefficient of the liquid beads to be the same as for a solid smooth sphere, $C_D=0.5$. For the detachment dynamics, its is important to compute the Weber number, who compares the inertia to capillarity: $We= \rho_\text{air}U_W^2H/\sigma$.  For most experiments $0<We<1.8$, for the detachment experiments, the Weber number reach up to $We=25$. In the film, the capillary forces play an important role,  which makes it therefore relevant to compare the length-scales with the capillary length $l_c=\sqrt{\sigma /(\rho g)}$. The ratio between the viscous forces and the capillary forces in the sliding liquid beads is then estimated through the capillary number $Ca=\mu u/\sigma$, which vary between $0.025<Ca<0.6$ in our experiments.

\section{Results}\label{results}

\subsection{Geometry of the flowing film}\label{effectshape}

\begin{figure}
\includegraphics[width=\textwidth]{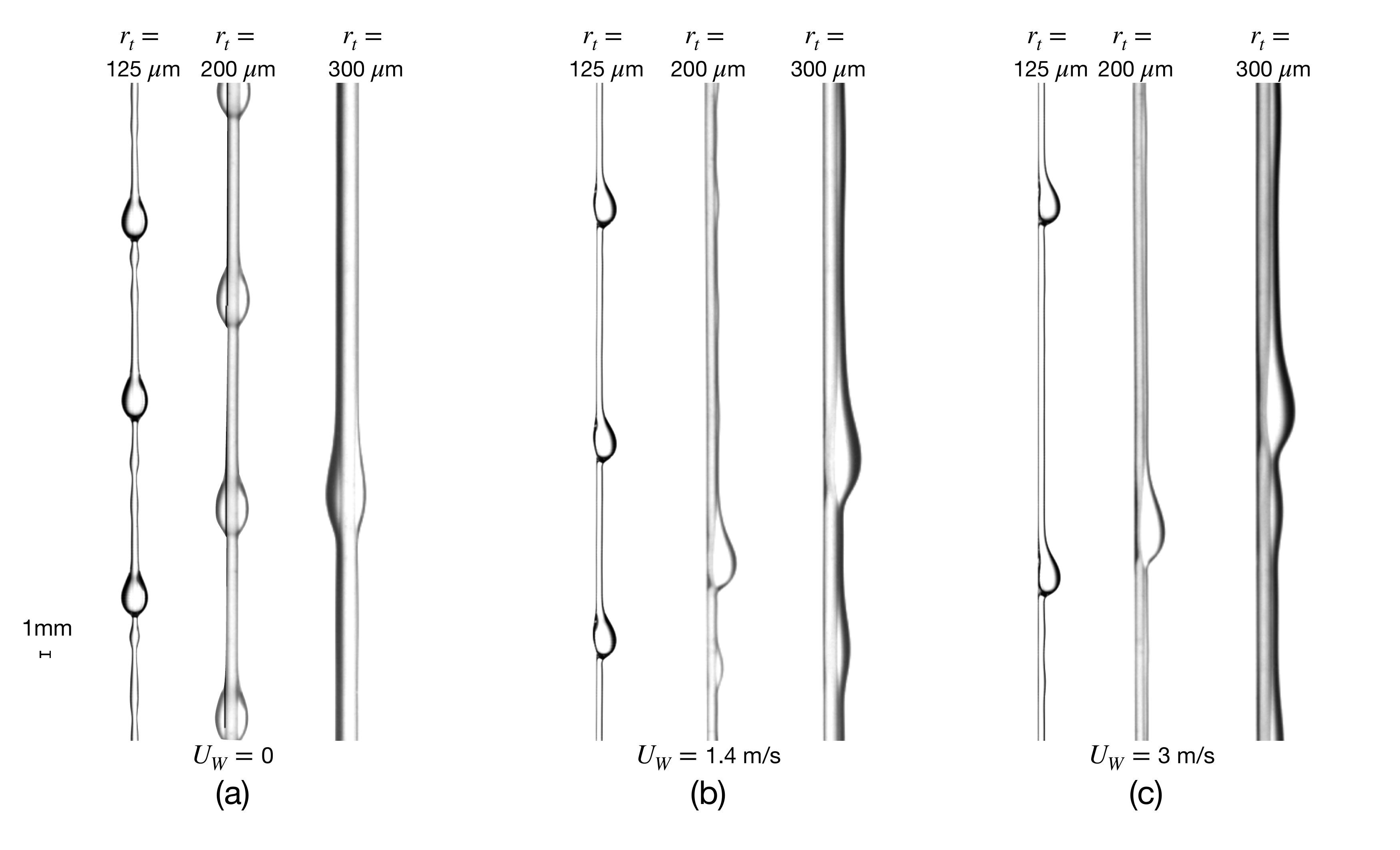}
\caption{Example of images of the film flowing down a thread with different radii, $r_t$. (a) Without wind. We see a regular and axisymmetric bead-pattern. (b) With a wind-speed $U_w=1.4$ m/s. The beads are pushed downstream (to the right of the pictures) and the distance between the beads are larger. (c) With a wind-speed $U_w=3$ m/s, for $r_t$=125 $\mu$m, the beads are pushed even further downstream and the distance between them further increased.}
\label{picture}
\end{figure}

 \begin{figure}
\includegraphics[width=0.9\textwidth]{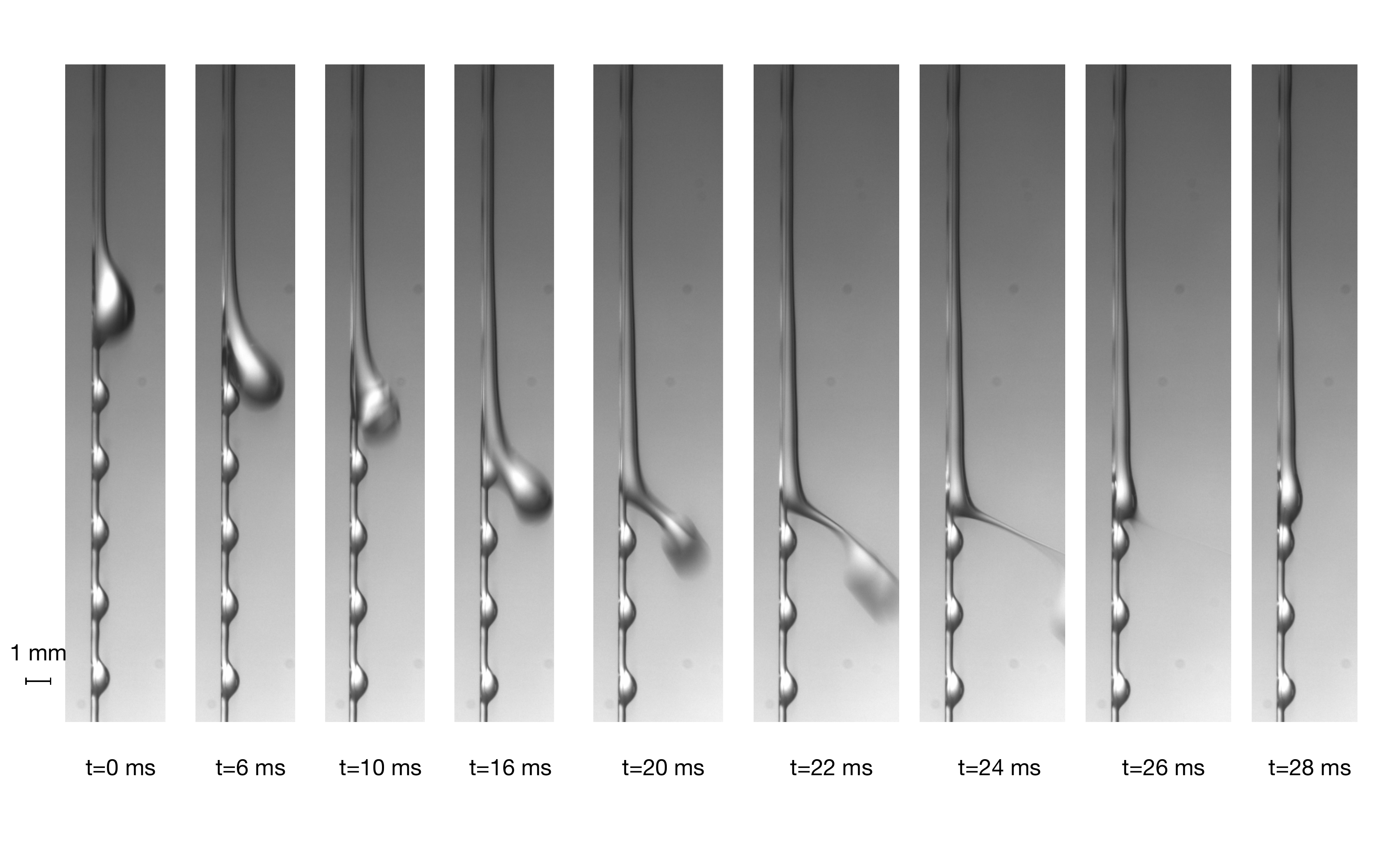}
\caption{Images of a bead detaching from the thread under the action of wind for $r_t=125$ $\mu$m and $U_W$=13 m/s.}
\label{detach}
\end{figure}
We place ourselves in the Rayleigh-Plateau regime, where the film pattern is regular, as already described by \cite{Duprat_2007}. Figure \ref{picture} shows images of the film shape around the thread for different diameters of the thread and wind speed. Without wind (Figure \ref{picture} (a)), the film destabilizes into a periodic, beading pattern. The liquid beads are axisymmetric of height $h = h_1=h_2$ (see Figure \ref{setup} (b)). This is true for all diameters of the thread. In the presence of wind, the liquid film is subjected to a drag force that pushes the liquid beads downstream. The axisymmetry is then broken and $h_1>h_2$ as we can observe in Figure \ref{picture} (b) and (c). In the Figure \ref{picture} (b), we can see that for intermediate wind speed, $h_2$ is already zero for $r_t=200$ $\mu$m and $r_t=300$ $\mu$m, when for $r_t=125$ $\mu$m, the bead still surrounds the fiber. Past a certain wind speed, all beads are completely on one side of the thread, $h_2=0$, and the shape of the beads does not depend on the wind speed, as we can see in Figure \ref{picture}  (c). The size of the beads is the same with or without wind. The distance between the beads, however, is larger with increasing wind speed. As we will discuss in more detail below, the beads slide faster when there is a side-wind, which is why at constant flow rate the distant between the beads increases, see Fig. \ref{picture}.
 
When the Reynolds number of the wind reaches a critical value $Re_{Wc}$, the liquid beads detach from the thread, as observed in Figure \ref{detach}. The dynamic of the detachment is complex. Under strong side-wind, the beads oscillate due to the vortex shedding of their wake. After some oscillations, a neck forms between the bead and the thread. The neck gets very thin and then breaks. Sahu et al. \cite{Sahu_2013} described the formation of a neck when a drop is detached from a thread by the action of a side-wind. 
 
\begin{figure}
\includegraphics[width=\textwidth]{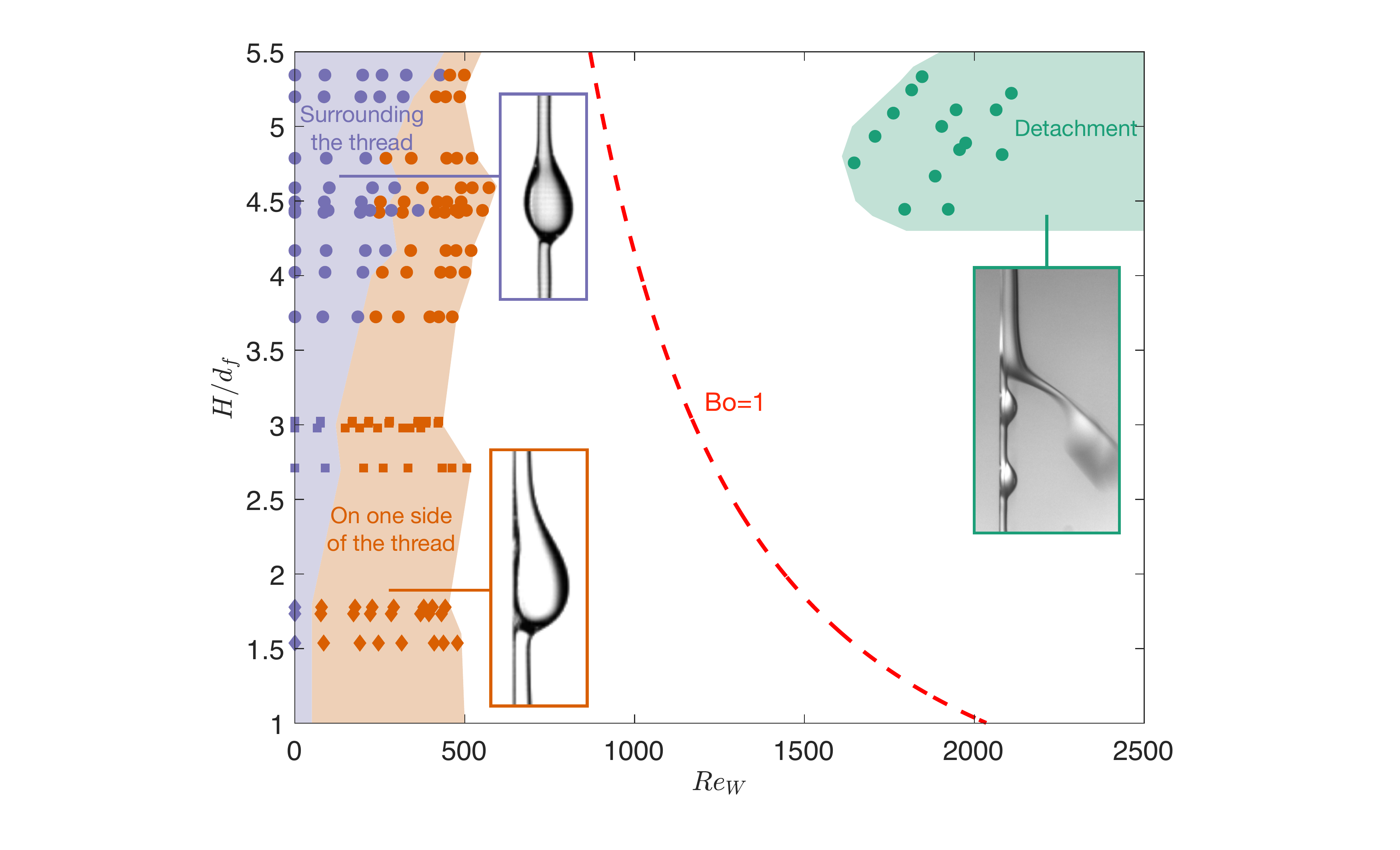}
\caption{Flow regime map for the position of the bead in the parameter space $Re_W$ and $H/d_f$. There are three regimes: (i) the beads surround the thread (in purple), (ii) the beads are pushed on one side of the thread (in orange), (iii) the beads detach from the thread (in green). The points represent experimental data for $r_t=$125 $\mu$m ($\bigcirc$), 200 $\mu$m ($\square$)  and 300 $\mu$m ($\Diamond$). The red dashed line correspond to $Bo=1$.}
\label{shape}
\end{figure}

 To understand the effect of side-wind we systematically change $Re_W$ and Figure \ref{shape} represents the regime map of the position of the liquid beads on the thread. The diagram is drawn in the plane $Re_W$ versus $H/d_f$. We have identified three main regions. In the first one, at low wind Reynolds number, the bead surrounds the thread (see the purple region in Fig. \ref{shape}). At larger Reynolds number, the liquid beads have a stationary shape (in their frame of reference) on one side of the thread (orange region in Fig. \ref{shape}). For $Re_W>Re_{Wc}$, the liquid beads detach. The dashed red line show when $Bo=1$, where the capillary force balance the drag force, which is the minimal requirement for detachment. In order to observe the detachment of the beads, we had to significantly reduce the working section of our experiments to increase the wind speed. The reduced observation window does not allow us to capture detachment for most experiments. The green points in Fig. \ref{shape} represent the experiments where we could see the detachment of the liquid beads. We only observed detachment for the smallest radius of the thread $r_t=125$ $\mu$m, because the capillary force increases with $r_t$. 
 

\subsection{Modification of the flow pattern by wind}

\begin{figure}
\includegraphics[width=\textwidth]{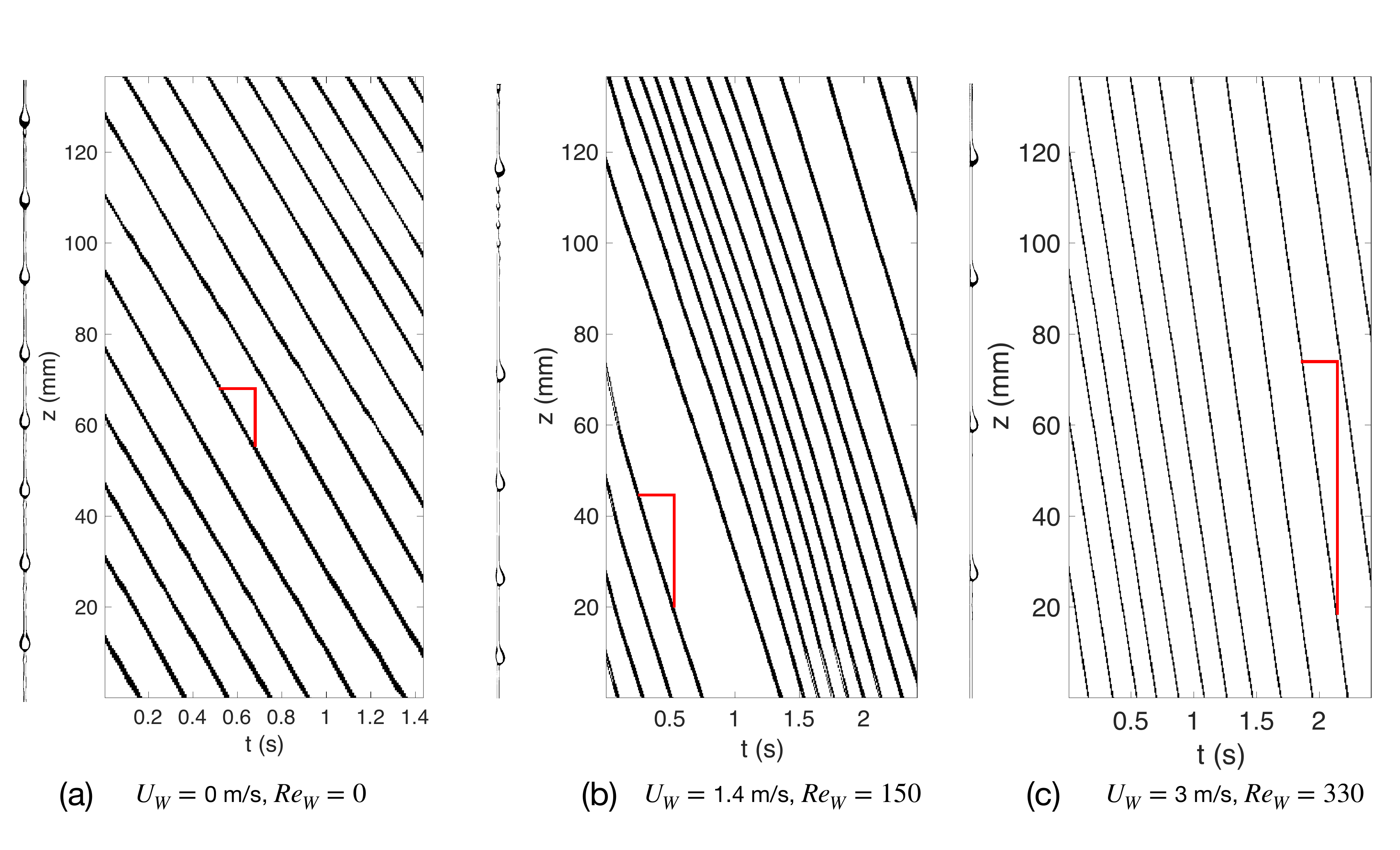}
\caption{Images of the flowing film and spatiotemporal diagrams of the flow pattern (a): without wind, (b): with a windspeed $U_\text{w}$=1.4 m/s, $Re_W=150$, (c): with a windspeed $U_\text{w}$=3 m/s, $Re_W=320$. The spatiotemporal diagrams are obtained by taking the temporal evolution of one vertical pixel line of the images of the film. The black lines correspond to the trace of the different beads.}
\label{st}
\end{figure}

To visualize the flow pattern, we draw the spatiotemporal diagram of the film in the following way. 
In Figure \ref{st} (a) on the left is shown a picture of the flowing film without wind for $r_t=125$ $\mu$m. The temporal evolution of a single vertical pixel line of this picture is taken to draw the spatiotemporal diagram of the flow, shown on the right of Figure \ref{st} (a). The black lines of the diagram represent the sliding beads. They are linear, parallel and regularly spaced, therefore the beads slide at constant velocity and the pattern is periodic. In the Figure \ref{st} (b) and (c), the spatiotemporal diagrams are drawn the same way for the same parameters, but for experiments with an air flow at a velocity (b): $U_W=1.4$ m/s ($Re_W=150$) and (c) $U_W=3$ m/s ($Re_W=320$). For intermediate wind speed the position of the bead on the thread is very sensitive to the small variation of the wind speed, this is why the bead pattern loses its regularity in Figure \ref{st} (b). The pattern becomes regular again when the bead is pushed completely on one side as is shown in Figure \ref{st} (c). We observe that the dark lines representing the beads are steeper with wind, which shows that the beads slide faster in the presence of wind. This faster sliding speed of the liquid beads in wind can be linked to the change of geometry of the film by the drag force.

\begin{figure}
\includegraphics[width=0.9\textwidth]{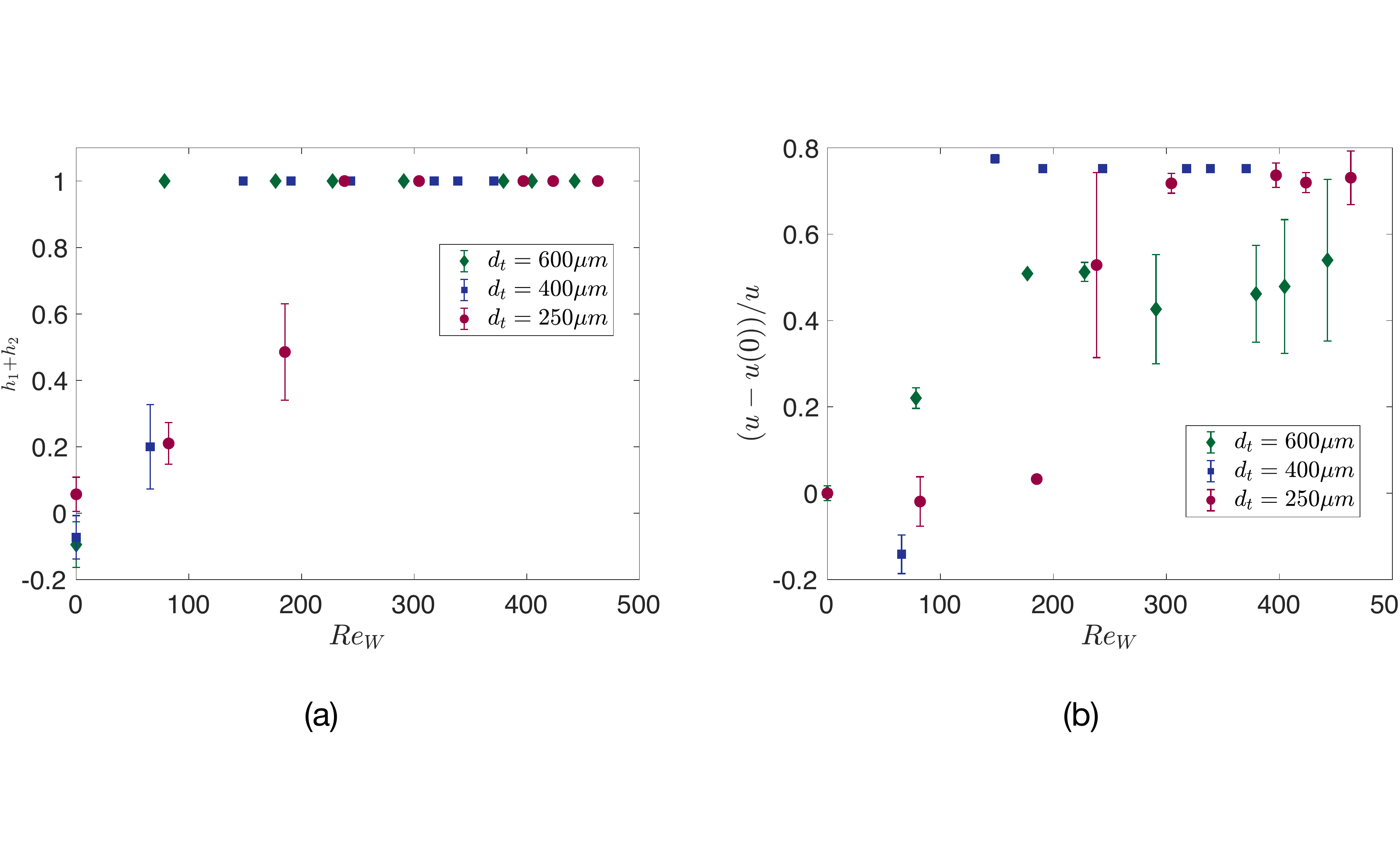}
\caption{(a) Bead asymmetry coefficient, $a$ as a function $Re_W$ for different values of $r_t$: $r_t=$125 $\mu$m ($\bigcirc$), 200 $\mu$m ($\square$)  and 300 $\mu$m ($\Diamond$). (b) Reduced velocity as a function of the Reynolds number of the wind flow, the parameters are the same as in (a).}
\label{uh}
\end{figure}

The change of bead geometry is quantified via $a=(h_1-h_2)/(h_1+h_2)$. As shown in Figure \ref{uh}, for all considered values of $r_t$, $a$ first increases monotonically with the Reynolds number $Re_W$ and then reaches the value one and stays constant. The liquid bead is then completely on one side of the thread and its shape does not change with wind speed. $a$ increases faster for larger $r_t$, see Figure \ref{picture}. In Figure \ref{uh} (b) shows the evolution of the Reynolds number of the reduced velocity of the bead $(u-u(0))/u(0)$, where $u(0)$ is the velocity without wind, for the same parameters as in (a). We can see that when $a$ reaches one, the velocity of the bead also arrives at a plateau and remains approximately constant when the Reynolds number is further increased. The increase of velocity seems to be linked to the loss of symmetry of the beads. For intermediate values of $a$, at lower $Re_W$ however, the velocity is not increased, and can even slightly decrease compared to the beads without wind. 

\begin{figure}
\includegraphics[width=0.7\textwidth]{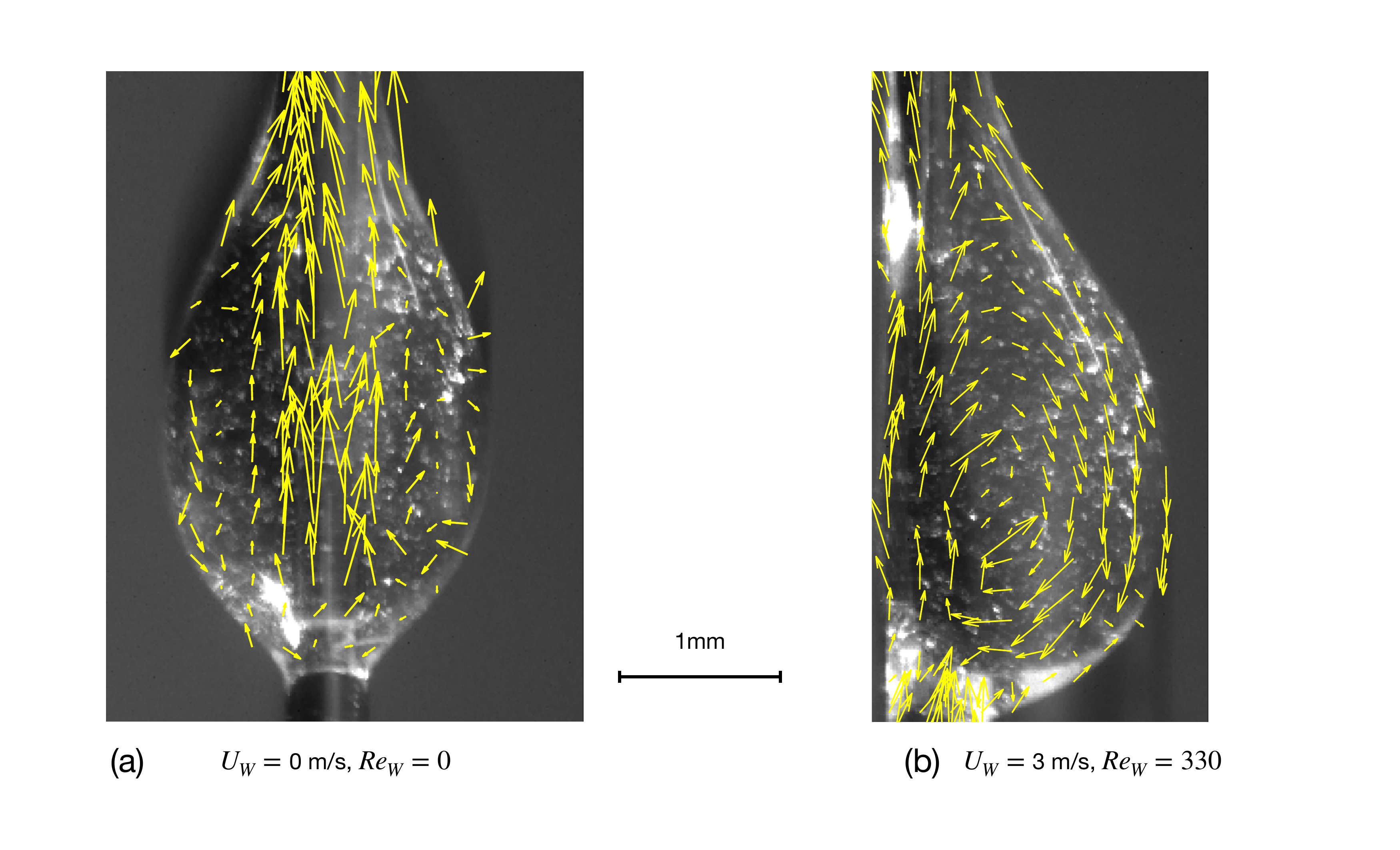}
\caption{Velocity field retrieved by PIV measurements of the flow inside the film in the frame of reference of the bead. (a) Without wind. (b) With a wind speed $U_W=3$ m/s, $Re_W=330$.}
\label{vect}
\end{figure}

The change in the liquid beads sliding speed can be explained by the modification of the shear in the liquid due to the change of the position of the beads relative to the thread, i. e. a modification of the viscous dissipation. We visualize the flow inside the beads through PIV measurement of the velocity field. Figure \ref{vect} shows the velocity fields inside the film in the frame of reference of the sliding bead for $r_t=125$ $\mu$m, (a) without wind and (b) with a wind speed $U_W=3$ m/s, $Re_W=330$. We note that the vector field is averaged over the thickness, $l_\text{laser}$, of the light sheet which is comparable to the size of the liquid bead ($l_\text{laser}\approx500$ $\mu$m). Therefore, it is quite difficult to obtain all details of the flow recirculation when there is no wind because of the axisymmetry. We can nonetheless observe in (a) that the flow is going upwards near the thread and seems to point downwards near the bead's interface. For similar parameters, Ruyer-Quil \textit{et al.} \cite{Ruyer-Quil_2008} shows in their numerical simulations, that there is an axisymmetric recirculation in the volume of the bead. Even though experimental limitations hinder us to observe all flow details of this recirculation without wind, our vector field is consistent with an axisymmetric recirculation in the volume of the bead. When the wind breaks the axisymmetry, the velocity field clearly shows a recirculation in the volume of the liquid bead (see Figure \ref{vect} (b)). 

The viscous dissipation inside the liquid film is given by the sum of different contributions: the dissipation at the edge (`apparent' contact angle) of the liquid beads, the dissipation in the thin film and the dissipation in the bulk of the beads. If we assume that the dissipation in the volume dominates, we can write the dissipation $\Phi_\mu$:
\begin{equation}
\Phi_\mu\sim \mu \int_\Omega \left(\frac{\partial v}{\partial r}\right)^2 d\Omega,
\label{diss1}
\end{equation}

assuming the variations in the vertical flow to predominantly be in the $r$-direction. $\Omega$ is the volume of the liquid bead and $v$ the velocity field of the liquid inside the bead.

When there is no wind and the system is axisymmetric, we can assume the scaling relation: 
${\partial v}/{\partial r}\sim u/h$, with $u$ the sliding velocity of the liquid bead and $h=h_1=h_2$, the height of the bead,
we get:
\begin{equation}
\Phi_\mu\sim \mu \Omega \left( \frac u {h}\right)^2.
\label{diss}
\end{equation}

\begin{figure}
\includegraphics[width=\textwidth]{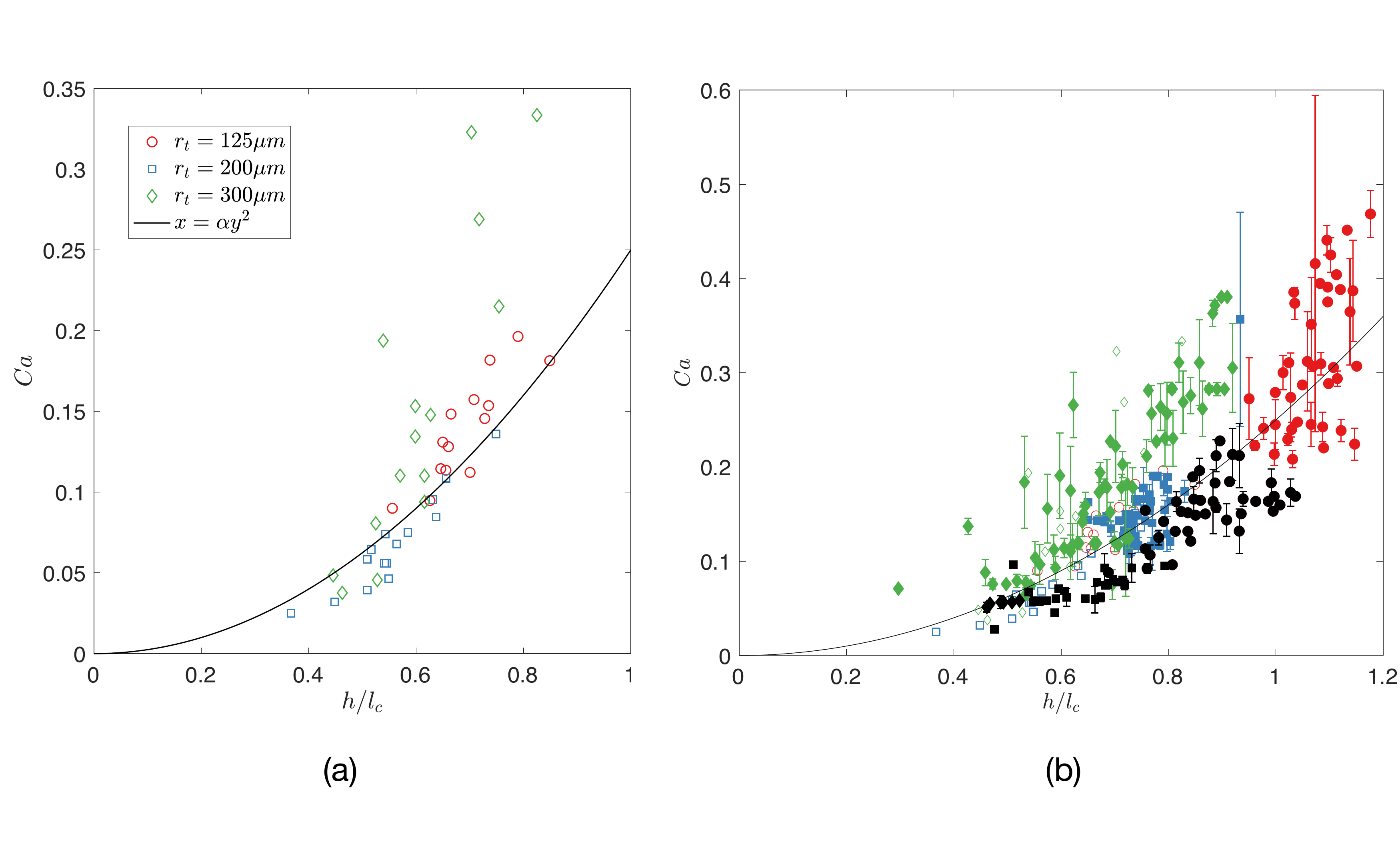}
\caption{(a) Capillary number as a function of the reduced height of the bead without wind. The points represent experimental data for $r_t=$125 $\mu$m ($\bigcirc$), 200 $\mu$m ($\square$)  and 300 $\mu$m ($\Diamond$). The black line is the theoretical prediction $x=\alpha y^2$ with $\alpha=0.25$. (b) Capillary number, $Ca=\mu u/\sigma$, as a function of the reduced downstream height of the bead without wind (empty symbols)and with wind (full symbols). The black symbols correspond to the experimental points where the liquid beads are not fully pushed on one side.}
\label{cfh}
\end{figure}

By balancing $\Phi_\mu$ with the variation of gravitational energy $\Phi_g=\rho g \Omega u$, we get a scaling relation for the sliding speed:

\begin{equation}
u \sim \frac {\rho g h^2} \mu.
\label{3}
\end{equation}
If we inject \ref{3} in the expression of the capillary number $Ca=\mu U /\sigma$, we get:

\begin{equation}
Ca= \alpha \left( \frac {h} {l_c} \right)^2,
\end{equation}

with $\alpha$ a scaling factor extracted from the experiments.
As shown in Figure \ref{cfh} (b), our experimental data follow a trend according to this scaling prediction when using $\alpha=0.25$  for $r_t=125$ and $r_t=200$ $\mu$m (respectively ($\bigcirc$) and ($\square$) in Figure \ref{cfh} (a)). For $r_t=300$ $\mu$m ($\Diamond$ in Figure \ref{cfh} (a)), as we can see in Figure \ref{picture} (a), the geometry is very different, as the liquid beads are approximately as thick as the film. The film must then be taken into account when calculating the viscous dissipation, giving one explanation to why the experimental points deviate some from the scaling prediction. 

With side-wind, when $h_1\gg h_2$, the volume of the upstream part of the drop is very small compared to the one of the downstream part. We then only  account for the contribution of the downstream part of the bead and we get then ${\partial v}/{\partial r}\sim u/h_1$, in the expression of the viscous dissipation. We get a similar expression for the capillary number as with wind:
\begin{equation}
Ca= \alpha \left( \frac {h_1} {l_c} \right)^2.
\label{ca}
\end{equation}
When $h_1\gtrsim h_2$, the contribution in the upstream part of the bead can not be neglected. There, the shear is higher because the height is smaller. This is why for intermediate values of the Reynolds number, we observe no increase or even a slight reduction of the sliding velocity of the bead in Figure \ref{uh} (b).

In Figure \ref{cfh} (b) we show $Ca$ as a function of $h_1/l_c$ for all data points without wind (empty symbols) and with wind (full symbols) for all considered wind-speed. The black symbols correspond to  $h_1\gtrsim h_2$. For $r_t=300$ $\mu$m, the experimental points are above the theoretical prediction (\ref{ca}). As without wind, the thick film plays a role that we did not take into account when calculating the viscous dissipation. For $r_t=125$ and 200 $\mu$m, the black points are as expected below the line  corresponding to (\ref{ca}). The other points, for $h_1\gg h_2$, are close to the theoretical line. The errorbars are nonetheless greater when there is wind, because as we saw on the spatio-temporal diagrams of Figure \ref{st}, the flow can also lose its regularity

\begin{figure}
\includegraphics[width=\textwidth]{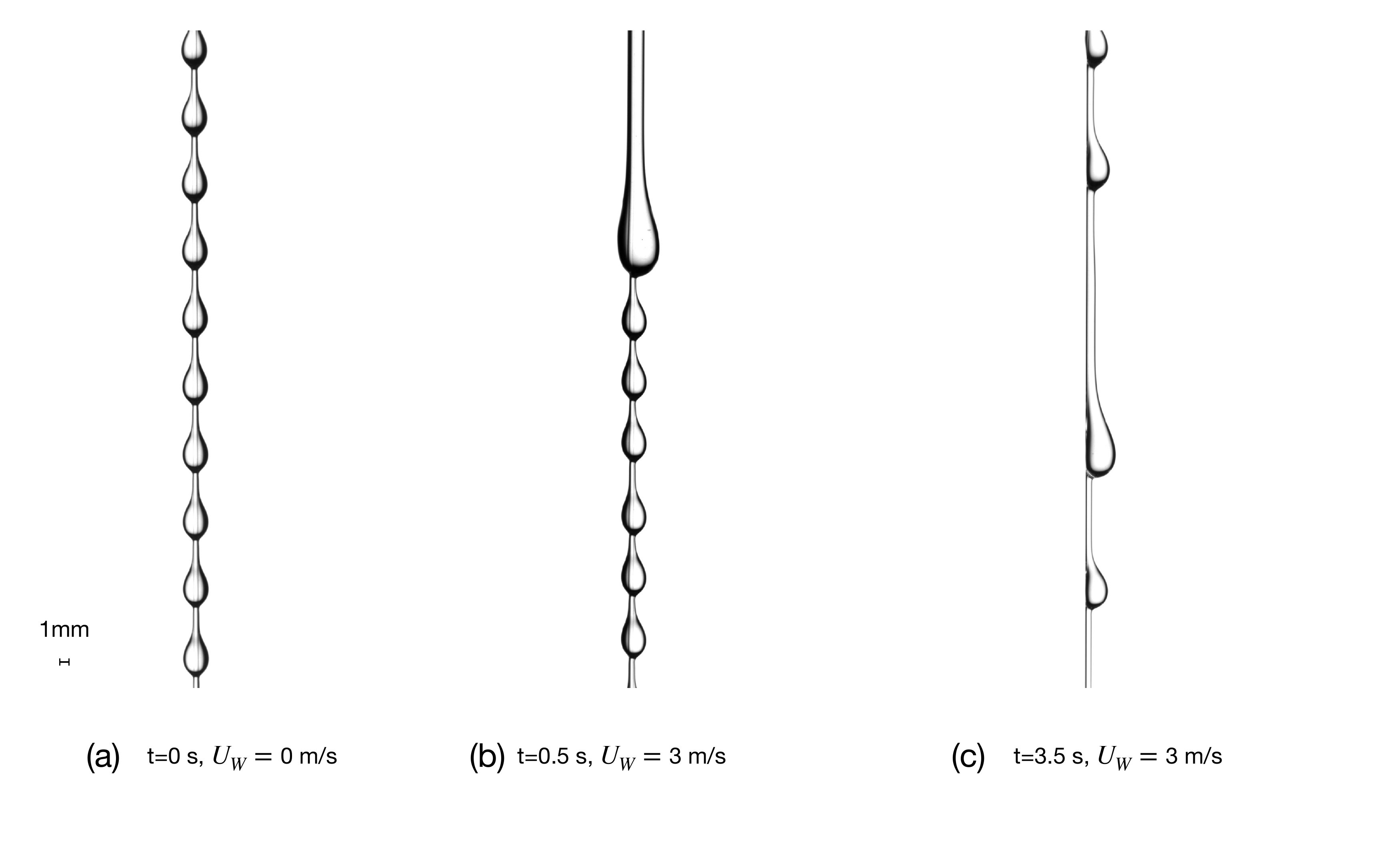}
\caption{Images of the thin film where the wind induce a change of flow regime. (a) Without any wind ($U_W=0$), defined as $t=0$. The beads are axisymmetric and regularly spaced. (b) Right after the fans have been switched on ($t>0$). A larger bead form and coalesce with the smaller beads, $t=0.5 $s. (c) With a wind-speed $U_W=3$ m/s, the flow pattern is not regular, we can see beads of different sizes irregularly spaced, $t=1.5 $s.}
\label{regime}
\end{figure}

We chose our parameters in order to be in the Rayleigh-Plateau regime where the beading pattern is regular without wind. However, as we have seen previously, the beads slide faster in the presence of wind. If the flow rate is large, the increase of inertia can lead to a change of flow regime \cite{Kliakhandler_2001}. Then larger beads appear and coalesce with small beads in an irregular manner. Pictures of such a change of regime are shown in Figure \ref{regime}. On (a) there is no wind, we observe small liquid beads regularly spaced. Then on (b), right after the wind flow has been turned on, the pattern becomes unstable and a large bead appears and coalesces with the smaller beads. The flow stays irregular, as is shown in (c), large beads appear and coalesce with smaller beads. The presence of a side-wind can therefore allow us to passively switch the regime of the film flow.

\section{Conculsion}\label{c}
In this article, we have investigated the effect of wind on a liquid film flowing down a thread in the Rayleigh-Plateau regime. We show that side-wind has an important non-trivial effect on the system as the drag force pushes the beads downstream, breaking the flow symmetry. We observe three different flow regimes depending on the wind Reynolds number: the beads surround the thread, the beads are pushed on one side of the thread or the beads detach. We show that this change of interfacial shape, as the bead is pushed on one side of the thread, leads to a decrease of the dissipation in the fluid. Consequently, the beads slide almost twice as fast in the presence of wind. By assuming that the dissipation is primarily in the volume of the bead, we obtain a scaling law for the slide velocity as a function of the height of the beads, which our experimental measurements follow. Finally, we see that side-wind, by actively increasing the slide velocity of the beads, can alter the flow regime of the liquid film and make it transition from a periodic to an irregular pattern. Side-wind allows us to control the film pattern and the wind induced increase in slide speed of the beads could be useful in draining systems and for droplet transport in nets.

Wind is present in many situations involving drops on threads and fibers, such as in fog harvesting systems used to collect water from fog in arid regions. The dynamics of how fog droplets are caught by fog-nets is highly complex and its understanding is primordial to the optimization of the water capture. 
We have shown here that side-wind can significantly influence the flow when an array of droplike structures slide down a thread. We can expect a similar effect also for partially wetting drops on threads or fibres, where wind would reduce the area of contact between the drop and the solid leading to an increase of the slide speed. The presence of a contact line adds richness to the system, where its dynamics will be affected by the drag force from the wind and is an interesting path for future studies. Gilet et al. \cite{Gilet_2010} studied how a drop sliding across a junction of threads will slow down and leave a small amount of liquid behind. These junctions are important when transporting drops on nets, where the drop dynamics can likely be influenced by the wind. In our experiments, the Nylon threads were strongly tightened to avoid wind-induced vibrations, whereas oscillations of the threads/net could generate additional effects that can influence the drop transport. Our work highlight the influence of wind on thin film flow on a thread, but there many other interesting avenues for future work of wind induced interfacial dynamics.

\section*{Acknowledgements}
We thank Olav Gundersen for his help with the experimental setup, Atle Jensen for his help with setup of the PIV system and Stéphane Poulain for fruitful discussions.

We acknowledge the financial support of the Research Council of Norway through the program NANO2021, project number 301138.




\begin{thebibliography}{29}%
\makeatletter
\providecommand \@ifxundefined [1]{%
 \@ifx{#1\undefined}
}%
\providecommand \@ifnum [1]{%
 \ifnum #1\expandafter \@firstoftwo
 \else \expandafter \@secondoftwo
 \fi
}%
\providecommand \@ifx [1]{%
 \ifx #1\expandafter \@firstoftwo
 \else \expandafter \@secondoftwo
 \fi
}%
\providecommand \natexlab [1]{#1}%
\providecommand \enquote  [1]{``#1''}%
\providecommand \bibnamefont  [1]{#1}%
\providecommand \bibfnamefont [1]{#1}%
\providecommand \citenamefont [1]{#1}%
\providecommand \href@noop [0]{\@secondoftwo}%
\providecommand \href [0]{\begingroup \@sanitize@url \@href}%
\providecommand \@href[1]{\@@startlink{#1}\@@href}%
\providecommand \@@href[1]{\endgroup#1\@@endlink}%
\providecommand \@sanitize@url [0]{\catcode `\\12\catcode `\$12\catcode
  `\&12\catcode `\#12\catcode `\^12\catcode `\_12\catcode `\%12\relax}%
\providecommand \@@startlink[1]{}%
\providecommand \@@endlink[0]{}%
\providecommand \url  [0]{\begingroup\@sanitize@url \@url }%
\providecommand \@url [1]{\endgroup\@href {#1}{\urlprefix }}%
\providecommand \urlprefix  [0]{URL }%
\providecommand \Eprint [0]{\href }%
\providecommand \doibase [0]{https://doi.org/}%
\providecommand \selectlanguage [0]{\@gobble}%
\providecommand \bibinfo  [0]{\@secondoftwo}%
\providecommand \bibfield  [0]{\@secondoftwo}%
\providecommand \translation [1]{[#1]}%
\providecommand \BibitemOpen [0]{}%
\providecommand \bibitemStop [0]{}%
\providecommand \bibitemNoStop [0]{.\EOS\space}%
\providecommand \EOS [0]{\spacefactor3000\relax}%
\providecommand \BibitemShut  [1]{\csname bibitem#1\endcsname}%
\let\auto@bib@innerbib\@empty
\bibitem [{\citenamefont {Kliakhandler}\ \emph {et~al.}(2001)\citenamefont
  {Kliakhandler}, \citenamefont {Davis},\ and\ \citenamefont
  {Bankoff}}]{Kliakhandler_2001}%
  \BibitemOpen
  \bibfield  {author} {\bibinfo {author} {\bibfnamefont {I.~L.}\ \bibnamefont
  {Kliakhandler}}, \bibinfo {author} {\bibfnamefont {S.~H.}\ \bibnamefont
  {Davis}},\ and\ \bibinfo {author} {\bibfnamefont {S.~G.}\ \bibnamefont
  {Bankoff}},\ }\bibfield  {title} {\bibinfo {title} {Viscous beads on vertical
  fibre},\ }\href@noop {} {\bibfield  {journal} {\bibinfo  {journal} {Journal
  of Fluid Mechanics}\ } (\bibinfo {year} {2001})}\BibitemShut {NoStop}%
\bibitem [{\citenamefont {Zeng}\ \emph {et~al.}(2017)\citenamefont {Zeng},
  \citenamefont {Sadeghpour}, \citenamefont {Warrier},\ and\ \citenamefont
  {Ju}}]{Zeng_2017}%
  \BibitemOpen
  \bibfield  {author} {\bibinfo {author} {\bibfnamefont {Z.}~\bibnamefont
  {Zeng}}, \bibinfo {author} {\bibfnamefont {A.}~\bibnamefont {Sadeghpour}},
  \bibinfo {author} {\bibfnamefont {G.~R.}\ \bibnamefont {Warrier}},\ and\
  \bibinfo {author} {\bibfnamefont {Y.~S.}\ \bibnamefont {Ju}},\ }\bibfield
  {title} {\bibinfo {title} {Experimental study of heat transfer between thin
  liquid films flowing down a vertical string in the rayleigh-plateau
  instability regime and a counterflowing gas stream},\ }\href@noop {}
  {\bibfield  {journal} {\bibinfo  {journal} {International Journal of Heat and
  Mass Transfer}\ } (\bibinfo {year} {2017})}\BibitemShut {NoStop}%
\bibitem [{\citenamefont {Chinju}\ \emph {et~al.}(2000)\citenamefont {Chinju},
  \citenamefont {Uchiyama},\ and\ \citenamefont {Mori}}]{Chinju_2000}%
  \BibitemOpen
  \bibfield  {author} {\bibinfo {author} {\bibfnamefont {H.}~\bibnamefont
  {Chinju}}, \bibinfo {author} {\bibfnamefont {K.}~\bibnamefont {Uchiyama}},\
  and\ \bibinfo {author} {\bibfnamefont {Y.~H.}\ \bibnamefont {Mori}},\
  }\bibfield  {title} {\bibinfo {title} {“string-of-beads” flow of liquids
  on vertical wires for gas absorption},\ }\href@noop {} {\bibfield  {journal}
  {\bibinfo  {journal} {AIChE Journal}\ }\textbf {\bibinfo {volume} {46}},\
  \bibinfo {pages} {937} (\bibinfo {year} {2000})}\BibitemShut {NoStop}%
\bibitem [{\citenamefont {Sadeghpour}\ \emph {et~al.}(2019)\citenamefont
  {Sadeghpour}, \citenamefont {Zeng}, \citenamefont {Ji}, \citenamefont
  {Ebrahimi}, \citenamefont {Bertozzi},\ and\ \citenamefont
  {Ju}}]{Sadeghpour_2019}%
  \BibitemOpen
  \bibfield  {author} {\bibinfo {author} {\bibfnamefont {A.}~\bibnamefont
  {Sadeghpour}}, \bibinfo {author} {\bibfnamefont {Z.}~\bibnamefont {Zeng}},
  \bibinfo {author} {\bibfnamefont {H.}~\bibnamefont {Ji}}, \bibinfo {author}
  {\bibfnamefont {N.~D.}\ \bibnamefont {Ebrahimi}}, \bibinfo {author}
  {\bibfnamefont {A.~L.}\ \bibnamefont {Bertozzi}},\ and\ \bibinfo {author}
  {\bibfnamefont {Y.~S.}\ \bibnamefont {Ju}},\ }\bibfield  {title} {\bibinfo
  {title} {Water vapor capturing using an array of traveling liquid beads for
  desalination and water treatment},\ }\href
  {https://www.science.org/doi/abs/10.1126/sciadv.aav7662} {\bibfield
  {journal} {\bibinfo  {journal} {Science Advances}\ }\textbf {\bibinfo
  {volume} {5}} (\bibinfo {year} {2019})}\BibitemShut {NoStop}%
\bibitem [{\citenamefont {Duprat}\ \emph {et~al.}(2009)\citenamefont {Duprat},
  \citenamefont {Ruyer-Quil},\ and\ \citenamefont
  {Giorgiutti-Dauphiné}}]{Duprat_2009}%
  \BibitemOpen
  \bibfield  {author} {\bibinfo {author} {\bibfnamefont {C.}~\bibnamefont
  {Duprat}}, \bibinfo {author} {\bibfnamefont {C.}~\bibnamefont {Ruyer-Quil}},\
  and\ \bibinfo {author} {\bibfnamefont {F.}~\bibnamefont
  {Giorgiutti-Dauphiné}},\ }\bibfield  {title} {\bibinfo {title} {Spatial
  evolution of a film flowing down a fiber},\ }\href@noop {} {\bibfield
  {journal} {\bibinfo  {journal} {Physics of Fluids}\ } (\bibinfo {year}
  {2009})}\BibitemShut {NoStop}%
\bibitem [{\citenamefont {Ruyer-Quil}\ \emph {et~al.}(2008)\citenamefont
  {Ruyer-Quil}, \citenamefont {Trevelyan}, \citenamefont {Treveleyan},
  \citenamefont {Giorgiutti-Dauphiné}, \citenamefont {Duprat},\ and\
  \citenamefont {Kalliadasis}}]{Ruyer-Quil_2008}%
  \BibitemOpen
  \bibfield  {author} {\bibinfo {author} {\bibfnamefont {C.}~\bibnamefont
  {Ruyer-Quil}}, \bibinfo {author} {\bibfnamefont {P.~M.~J.}\ \bibnamefont
  {Trevelyan}}, \bibinfo {author} {\bibfnamefont {P.}~\bibnamefont
  {Treveleyan}}, \bibinfo {author} {\bibfnamefont {F.}~\bibnamefont
  {Giorgiutti-Dauphiné}}, \bibinfo {author} {\bibfnamefont {C.}~\bibnamefont
  {Duprat}},\ and\ \bibinfo {author} {\bibfnamefont {S.}~\bibnamefont
  {Kalliadasis}},\ }\bibfield  {title} {\bibinfo {title} {Modelling film flows
  down a fibre},\ }\href@noop {} {\bibfield  {journal} {\bibinfo  {journal}
  {Journal of Fluid Mechanics}\ } (\bibinfo {year} {2008})}\BibitemShut
  {NoStop}%
\bibitem [{\citenamefont {Quéré}(1990)}]{Quere_1990}%
  \BibitemOpen
  \bibfield  {author} {\bibinfo {author} {\bibfnamefont {D.}~\bibnamefont
  {Quéré}},\ }\bibfield  {title} {\bibinfo {title} {Thin films flowing on
  vertical fibers},\ }\href@noop {} {\bibfield  {journal} {\bibinfo  {journal}
  {EPL}\ } (\bibinfo {year} {1990})}\BibitemShut {NoStop}%
\bibitem [{\citenamefont {Frenkel}(1992)}]{Frenkel_1992}%
  \BibitemOpen
  \bibfield  {author} {\bibinfo {author} {\bibfnamefont {A.~L.}\ \bibnamefont
  {Frenkel}},\ }\bibfield  {title} {\bibinfo {title} {Nonlinear theory of
  strongly undulating thin films flowing down vertical cylinders},\ }\href@noop
  {} {\bibfield  {journal} {\bibinfo  {journal} {EPL}\ } (\bibinfo {year}
  {1992})}\BibitemShut {NoStop}%
\bibitem [{\citenamefont {Kalliadasis}\ and\ \citenamefont
  {Chang}(1994)}]{Kalliadasis_1994}%
  \BibitemOpen
  \bibfield  {author} {\bibinfo {author} {\bibfnamefont {S.}~\bibnamefont
  {Kalliadasis}}\ and\ \bibinfo {author} {\bibfnamefont {H.-C.}\ \bibnamefont
  {Chang}},\ }\bibfield  {title} {\bibinfo {title} {Drop formation during
  coating of vertical fibres},\ }\href@noop {} {\bibfield  {journal} {\bibinfo
  {journal} {Journal of Fluid Mechanics}\ } (\bibinfo {year}
  {1994})}\BibitemShut {NoStop}%
\bibitem [{\citenamefont {Plateau}(1873)}]{Plateau}%
  \BibitemOpen
  \bibfield  {author} {\bibinfo {author} {\bibfnamefont {J.}~\bibnamefont
  {Plateau}},\ }\href@noop {} {\emph {\bibinfo {title} {Statique Expérimentale
  et Théorique des Liquides Soumis aux seules Forces Moléculaires}}}\
  (\bibinfo  {publisher} {Gauthier Villars Paris},\ \bibinfo {year}
  {1873})\BibitemShut {NoStop}%
\bibitem [{\citenamefont {Rayleigh}(1878)}]{Rayleigh}%
  \BibitemOpen
  \bibfield  {author} {\bibinfo {author} {\bibfnamefont {J.~W.}\ \bibnamefont
  {Rayleigh}},\ }\bibfield  {title} {\bibinfo {title} {On the instability of
  jets},\ }\href@noop {} {\bibfield  {journal} {\bibinfo  {journal}
  {Proceedings of the London Mathematical Society}\ }\textbf {\bibinfo {volume}
  {s1-10}},\ \bibinfo {pages} {4} (\bibinfo {year} {1878})}\BibitemShut
  {NoStop}%
\bibitem [{\citenamefont {Kapitza}(1965)}]{Kapitza}%
  \BibitemOpen
  \bibfield  {author} {\bibinfo {author} {\bibfnamefont {P.~L.}\ \bibnamefont
  {Kapitza}},\ }\href@noop {} {{\selectlanguage {und}\bibinfo {title}
  {Collected papers of p. l. kapitza : 2 : 1938-1964}}} (\bibinfo {year}
  {1965})\BibitemShut {NoStop}%
\bibitem [{\citenamefont {Sadeghpour}\ \emph {et~al.}(2017)\citenamefont
  {Sadeghpour}, \citenamefont {Zeng},\ and\ \citenamefont
  {Ju}}]{Sadeghpour_2017}%
  \BibitemOpen
  \bibfield  {author} {\bibinfo {author} {\bibfnamefont {A.}~\bibnamefont
  {Sadeghpour}}, \bibinfo {author} {\bibfnamefont {Z.}~\bibnamefont {Zeng}},\
  and\ \bibinfo {author} {\bibfnamefont {Y.~S.}\ \bibnamefont {Ju}},\
  }\bibfield  {title} {\bibinfo {title} {Effects of nozzle geometry on the
  fluid dynamics of thin liquid films flowing down vertical strings in the
  rayleigh-plateau regime.},\ }\href@noop {} {\bibfield  {journal} {\bibinfo
  {journal} {Langmuir}\ } (\bibinfo {year} {2017})}\BibitemShut {NoStop}%
\bibitem [{\citenamefont {Ji}\ \emph {et~al.}(2019)\citenamefont {Ji},
  \citenamefont {Falcon}, \citenamefont {Sadeghpour}, \citenamefont {Zeng},
  \citenamefont {Ju},\ and\ \citenamefont {Bertozzi}}]{Ji_2019}%
  \BibitemOpen
  \bibfield  {author} {\bibinfo {author} {\bibfnamefont {H.}~\bibnamefont
  {Ji}}, \bibinfo {author} {\bibfnamefont {C.}~\bibnamefont {Falcon}}, \bibinfo
  {author} {\bibfnamefont {A.}~\bibnamefont {Sadeghpour}}, \bibinfo {author}
  {\bibfnamefont {Z.}~\bibnamefont {Zeng}}, \bibinfo {author} {\bibfnamefont
  {Y.~S.}\ \bibnamefont {Ju}},\ and\ \bibinfo {author} {\bibfnamefont {A.~L.}\
  \bibnamefont {Bertozzi}},\ }\bibfield  {title} {\bibinfo {title} {Dynamics of
  thin liquid films on vertical cylindrical fibres},\ }\href@noop {} {\bibfield
   {journal} {\bibinfo  {journal} {Journal of Fluid Mechanics}\ } (\bibinfo
  {year} {2019})}\BibitemShut {NoStop}%
\bibitem [{\citenamefont {Ding}\ \emph {et~al.}(2014)\citenamefont {Ding},
  \citenamefont {Xie}, \citenamefont {Xie}, \citenamefont {Wong},\ and\
  \citenamefont {Liu}}]{Ding_2014}%
  \BibitemOpen
  \bibfield  {author} {\bibinfo {author} {\bibfnamefont {Z.}~\bibnamefont
  {Ding}}, \bibinfo {author} {\bibfnamefont {J.}~\bibnamefont {Xie}}, \bibinfo
  {author} {\bibfnamefont {J.}~\bibnamefont {Xie}}, \bibinfo {author}
  {\bibfnamefont {T.~N.}\ \bibnamefont {Wong}},\ and\ \bibinfo {author}
  {\bibfnamefont {R.}~\bibnamefont {Liu}},\ }\bibfield  {title} {\bibinfo
  {title} {Dynamics of liquid films on vertical fibres in a radial electric
  field},\ }\href@noop {} {\bibfield  {journal} {\bibinfo  {journal} {Journal
  of Fluid Mechanics}\ } (\bibinfo {year} {2014})}\BibitemShut {NoStop}%
\bibitem [{\citenamefont {Kabova}\ \emph {et~al.}(2012)\citenamefont {Kabova},
  \citenamefont {Kuznetsov},\ and\ \citenamefont {Kabov}}]{Kabova_2012}%
  \BibitemOpen
  \bibfield  {author} {\bibinfo {author} {\bibfnamefont {Y.}~\bibnamefont
  {Kabova}}, \bibinfo {author} {\bibfnamefont {V.}~\bibnamefont {Kuznetsov}},\
  and\ \bibinfo {author} {\bibfnamefont {O.}~\bibnamefont {Kabov}},\ }\bibfield
   {title} {\bibinfo {title} {Temperature dependent viscosity and surface
  tension effects on deformations of non-isothermal falling liquid film},\
  }\href {https://www.sciencedirect.com/science/article/pii/S0017931011005175}
  {\bibfield  {journal} {\bibinfo  {journal} {International Journal of Heat and
  Mass Transfer}\ }\textbf {\bibinfo {volume} {55}},\ \bibinfo {pages} {1271}
  (\bibinfo {year} {2012})}\BibitemShut {NoStop}%
\bibitem [{\citenamefont {Liu}\ \emph {et~al.}(2018)\citenamefont {Liu},
  \citenamefont {Chen}, \citenamefont {Chen},\ and\ \citenamefont
  {Ding}}]{Liu_2018}%
  \BibitemOpen
  \bibfield  {author} {\bibinfo {author} {\bibfnamefont {R.}~\bibnamefont
  {Liu}}, \bibinfo {author} {\bibfnamefont {X.}~\bibnamefont {Chen}}, \bibinfo
  {author} {\bibfnamefont {X.}~\bibnamefont {Chen}},\ and\ \bibinfo {author}
  {\bibfnamefont {Z.}~\bibnamefont {Ding}},\ }\bibfield  {title} {\bibinfo
  {title} {Absolute and convective instabilities of a film flow down a vertical
  fiber subjected to a radial electric field.},\ }\href@noop {} {\bibfield
  {journal} {\bibinfo  {journal} {Physical Review E}\ } (\bibinfo {year}
  {2018})}\BibitemShut {NoStop}%
\bibitem [{\citenamefont {Ding}\ \emph {et~al.}(2018)\citenamefont {Ding},
  \citenamefont {Liu}, \citenamefont {Wong},\ and\ \citenamefont
  {Yang}}]{Ding_2018}%
  \BibitemOpen
  \bibfield  {author} {\bibinfo {author} {\bibfnamefont {Z.}~\bibnamefont
  {Ding}}, \bibinfo {author} {\bibfnamefont {R.}~\bibnamefont {Liu}}, \bibinfo
  {author} {\bibfnamefont {T.~N.}\ \bibnamefont {Wong}},\ and\ \bibinfo
  {author} {\bibfnamefont {C.}~\bibnamefont {Yang}},\ }\bibfield  {title}
  {\bibinfo {title} {Absolute instability induced by marangoni effect in thin
  liquid film flows on vertical cylindrical surfaces},\ }\href@noop {}
  {\bibfield  {journal} {\bibinfo  {journal} {Chemical Engineering Science}\ }
  (\bibinfo {year} {2018})}\BibitemShut {NoStop}%
\bibitem [{\citenamefont {Liu}\ \emph {et~al.}(2020)\citenamefont {Liu},
  \citenamefont {Liu}, \citenamefont {Liu},\ and\ \citenamefont
  {Ding}}]{Liu_2020}%
  \BibitemOpen
  \bibfield  {author} {\bibinfo {author} {\bibfnamefont {R.}~\bibnamefont
  {Liu}}, \bibinfo {author} {\bibfnamefont {R.}~\bibnamefont {Liu}}, \bibinfo
  {author} {\bibfnamefont {R.}~\bibnamefont {Liu}},\ and\ \bibinfo {author}
  {\bibfnamefont {Z.}~\bibnamefont {Ding}},\ }\bibfield  {title} {\bibinfo
  {title} {Instabilities and bifurcations of liquid films flowing down a
  rotating fibre},\ }\href@noop {} {\bibfield  {journal} {\bibinfo  {journal}
  {Journal of Fluid Mechanics}\ } (\bibinfo {year} {2020})}\BibitemShut
  {NoStop}%
\bibitem [{\citenamefont {Boulogne}\ \emph {et~al.}(2013)\citenamefont
  {Boulogne}, \citenamefont {Fardin}, \citenamefont {Lerouge}, \citenamefont
  {Pauchard},\ and\ \citenamefont {Giorgiutti-Dauphiné}}]{Boulogne_SM_2013}%
  \BibitemOpen
  \bibfield  {author} {\bibinfo {author} {\bibfnamefont {F.}~\bibnamefont
  {Boulogne}}, \bibinfo {author} {\bibfnamefont {M.~A.}\ \bibnamefont
  {Fardin}}, \bibinfo {author} {\bibfnamefont {S.}~\bibnamefont {Lerouge}},
  \bibinfo {author} {\bibfnamefont {L.}~\bibnamefont {Pauchard}},\ and\
  \bibinfo {author} {\bibfnamefont {F.}~\bibnamefont {Giorgiutti-Dauphiné}},\
  }\bibfield  {title} {\bibinfo {title} {Suppression of the rayleigh–plateau
  instability on a vertical fibre coated with wormlike micelle solutions},\
  }\href@noop {} {\bibfield  {journal} {\bibinfo  {journal} {Soft Matter}\
  }\textbf {\bibinfo {volume} {9}},\ \bibinfo {pages} {7787} (\bibinfo {year}
  {2013})}\BibitemShut {NoStop}%
\bibitem [{\citenamefont {Boulogne}\ \emph {et~al.}(2012)\citenamefont
  {Boulogne}, \citenamefont {Pauchard},\ and\ \citenamefont
  {Giorgiutti-Dauphiné}}]{Boulogne_2012}%
  \BibitemOpen
  \bibfield  {author} {\bibinfo {author} {\bibfnamefont {F.}~\bibnamefont
  {Boulogne}}, \bibinfo {author} {\bibfnamefont {L.}~\bibnamefont {Pauchard}},\
  and\ \bibinfo {author} {\bibfnamefont {F.}~\bibnamefont
  {Giorgiutti-Dauphiné}},\ }\bibfield  {title} {\bibinfo {title} {Instability
  and morphology of polymer solutions coating a fibre},\ }\href@noop {}
  {\bibfield  {journal} {\bibinfo  {journal} {Journal of Fluid Mechanics}\
  }\textbf {\bibinfo {volume} {704}},\ \bibinfo {pages} {232–250} (\bibinfo
  {year} {2012})}\BibitemShut {NoStop}%
\bibitem [{\citenamefont {Gabbard}\ and\ \citenamefont
  {Bostwick}(2021)}]{Gabbard_2021}%
  \BibitemOpen
  \bibfield  {author} {\bibinfo {author} {\bibfnamefont {C.~T.}\ \bibnamefont
  {Gabbard}}\ and\ \bibinfo {author} {\bibfnamefont {J.}~\bibnamefont
  {Bostwick}},\ }\bibfield  {title} {\bibinfo {title} {Scaling analysis of the
  plateau–rayleigh instability in thin film flow down a fiber},\ }\href@noop
  {} {\bibfield  {journal} {\bibinfo  {journal} {Experiments in Fluids}\ }
  (\bibinfo {year} {2021})}\BibitemShut {NoStop}%
\bibitem [{\citenamefont {Gilet}\ \emph {et~al.}(2010)\citenamefont {Gilet},
  \citenamefont {Terwagne},\ and\ \citenamefont {Vandewalle}}]{Gilet_2010}%
  \BibitemOpen
  \bibfield  {author} {\bibinfo {author} {\bibfnamefont {T.}~\bibnamefont
  {Gilet}}, \bibinfo {author} {\bibfnamefont {D.}~\bibnamefont {Terwagne}},\
  and\ \bibinfo {author} {\bibfnamefont {N.}~\bibnamefont {Vandewalle}},\
  }\bibfield  {title} {\bibinfo {title} {Droplets sliding on fibres},\
  }\href@noop {} {\bibfield  {journal} {\bibinfo  {journal} {European Physical
  Journal E}\ } (\bibinfo {year} {2010})}\BibitemShut {NoStop}%
\bibitem [{\citenamefont {Mullins}\ \emph {et~al.}(2005)\citenamefont
  {Mullins}, \citenamefont {Braddock}, \citenamefont {Agranovski},
  \citenamefont {Cropp},\ and\ \citenamefont {O'Leary}}]{Mullins_2005}%
  \BibitemOpen
  \bibfield  {author} {\bibinfo {author} {\bibfnamefont {B.~J.}\ \bibnamefont
  {Mullins}}, \bibinfo {author} {\bibfnamefont {R.~D.}\ \bibnamefont
  {Braddock}}, \bibinfo {author} {\bibfnamefont {I.~E.}\ \bibnamefont
  {Agranovski}}, \bibinfo {author} {\bibfnamefont {R.~A.}\ \bibnamefont
  {Cropp}},\ and\ \bibinfo {author} {\bibfnamefont {R.~A.}\ \bibnamefont
  {O'Leary}},\ }\bibfield  {title} {\bibinfo {title} {Observation and modelling
  of clamshell droplets on vertical fibres subjected to gravitational and drag
  forces},\ }\href
  {https://www.sciencedirect.com/science/article/pii/S0021979704010331}
  {\bibfield  {journal} {\bibinfo  {journal} {Journal of Colloid and Interface
  Science}\ }\textbf {\bibinfo {volume} {284}},\ \bibinfo {pages} {245}
  (\bibinfo {year} {2005})}\BibitemShut {NoStop}%
\bibitem [{\citenamefont {Mullins}\ \emph {et~al.}(2006)\citenamefont
  {Mullins}, \citenamefont {Braddock}, \citenamefont {Agranovski},\ and\
  \citenamefont {Cropp}}]{Mullins_2006}%
  \BibitemOpen
  \bibfield  {author} {\bibinfo {author} {\bibfnamefont {B.}~\bibnamefont
  {Mullins}}, \bibinfo {author} {\bibfnamefont {R.}~\bibnamefont {Braddock}},
  \bibinfo {author} {\bibfnamefont {I.}~\bibnamefont {Agranovski}},\ and\
  \bibinfo {author} {\bibfnamefont {R.}~\bibnamefont {Cropp}},\ }\bibfield
  {title} {\bibinfo {title} {Observation and modelling of barrel droplets on
  vertical fibres subjected to gravitational and drag forces.},\ }\href@noop {}
  {\bibfield  {journal} {\bibinfo  {journal} {Journal of colloid and interface
  science}\ } (\bibinfo {year} {2006})}\BibitemShut {NoStop}%
\bibitem [{\citenamefont {Bintein}\ \emph {et~al.}(2019)\citenamefont
  {Bintein}, \citenamefont {Bense}, \citenamefont {Clanet},\ and\ \citenamefont
  {Quéré}}]{Bintein_2019}%
  \BibitemOpen
  \bibfield  {author} {\bibinfo {author} {\bibfnamefont {P.-B.}\ \bibnamefont
  {Bintein}}, \bibinfo {author} {\bibfnamefont {H.}~\bibnamefont {Bense}},
  \bibinfo {author} {\bibfnamefont {C.}~\bibnamefont {Clanet}},\ and\ \bibinfo
  {author} {\bibfnamefont {D.}~\bibnamefont {Quéré}},\ }\bibfield  {title}
  {\bibinfo {title} {Self-propelling droplets on fibres subject to a
  crosswind},\ }\href@noop {} {\bibfield  {journal} {\bibinfo  {journal}
  {Nature Physics}\ } (\bibinfo {year} {2019})}\BibitemShut {NoStop}%
\bibitem [{\citenamefont {Sahu}\ \emph {et~al.}(2013)\citenamefont {Sahu},
  \citenamefont {Sinha-Ray}, \citenamefont {Yarin},\ and\ \citenamefont
  {Pourdeyhimi}}]{Sahu_2013}%
  \BibitemOpen
  \bibfield  {author} {\bibinfo {author} {\bibfnamefont {R.~P.}\ \bibnamefont
  {Sahu}}, \bibinfo {author} {\bibfnamefont {S.}~\bibnamefont {Sinha-Ray}},
  \bibinfo {author} {\bibfnamefont {A.}~\bibnamefont {Yarin}},\ and\ \bibinfo
  {author} {\bibfnamefont {B.}~\bibnamefont {Pourdeyhimi}},\ }\bibfield
  {title} {\bibinfo {title} {Blowing drops off a filament},\ }\href@noop {}
  {\bibfield  {journal} {\bibinfo  {journal} {Soft Matter}\ } (\bibinfo {year}
  {2013})}\BibitemShut {NoStop}%
\bibitem [{\citenamefont {Thielicke}\ and\ \citenamefont
  {Sonntag}(2021)}]{PIVlab}%
  \BibitemOpen
  \bibfield  {author} {\bibinfo {author} {\bibfnamefont {W.}~\bibnamefont
  {Thielicke}}\ and\ \bibinfo {author} {\bibfnamefont {R.}~\bibnamefont
  {Sonntag}},\ }\bibfield  {title} {\bibinfo {title} {Particle image
  velocimetry for {MATLAB}: Accuracy and enhanced algorithms in {PIVlab}},\
  }\href {https://doi.org/10.5334%2Fjors.334} {\bibfield  {journal} {\bibinfo
  {journal} {Journal of Open Research Software}\ }\textbf {\bibinfo {volume}
  {9}} (\bibinfo {year} {2021})}\BibitemShut {NoStop}%
\bibitem [{\citenamefont {Duprat}\ \emph {et~al.}(2007)\citenamefont {Duprat},
  \citenamefont {Ruyer-Quil}, \citenamefont {Kalliadasis},\ and\ \citenamefont
  {Giorgiutti-Dauphiné}}]{Duprat_2007}%
  \BibitemOpen
  \bibfield  {author} {\bibinfo {author} {\bibfnamefont {C.}~\bibnamefont
  {Duprat}}, \bibinfo {author} {\bibfnamefont {C.}~\bibnamefont {Ruyer-Quil}},
  \bibinfo {author} {\bibfnamefont {S.}~\bibnamefont {Kalliadasis}},\ and\
  \bibinfo {author} {\bibfnamefont {F.}~\bibnamefont {Giorgiutti-Dauphiné}},\
  }\bibfield  {title} {\bibinfo {title} {Absolute and convective instabilities
  of a viscous film flowing down a vertical fiber},\ }\href@noop {} {\bibfield
  {journal} {\bibinfo  {journal} {Physical Review Letters}\ } (\bibinfo {year}
  {2007})}\BibitemShut {NoStop}%
\end{thebibliography}%
%

\end{document}